# 'Psychotic' Markov Blankets:
# Striking a Free Energy Balance for Complex Adaptation

Inês Hipólito [1,2]

1. Humboldt-Universität zu Berlin, Berlin School of Mind and Brain, Germany
2. Macquarie University, Philosophy Department, Centre for Agency, Values and Ethics

# Abstract

This paper proposes a framework for optimising the adaptation and attunement of a Complex Adaptive System (CAS) with its environments. The tendency towards stability can be explained by minimising free energy but high variability, noise, and over-specialized rigidity can lead to a "stuck state" in a CAS. Without perturbation (increasing free energy), the system remains stuck and unable to adapt to changing circumstances. The paper introduces the concept of 'psychotic' Markov blankets to understand and specify factors contributing to maladjustment conditions moving away from the minimum stuck state. The paper offers directions for optimising adaptation and attunement to be applied to real-world problems, from cells to behaviour, societies and ecosystems.

## Introduction

Living beings are complex, dynamical systems (CAS) that exhibit remarkable resilience and tenacity in the face of environmental challenges. While, by the second law of thermodynamics, open systems should tend towards dissipation and disorder, living systems appear to maintain and perpetuate themselves with precision and intentionality. Adapting to an ever-changing world is a generative process that requires not only dynamically interacting with the world, but interacting with a purpose: context-specific interaction as optimising adaptation to avoid the increasing entropy (H-theorem). In order to adapt to changing environments, systems must engage in context-specific interactions that develop towards the optimisation of maintaining health and well-being.

Complex Adaptive Systems (CAS) are systems that exhibit emergent behaviour arising from the interactions among their individual components. These systems are characterised by a large number of agents, which interact with one another and with their environment in a nonlinear and dynamic manner. A CAS is often studied in a variety of fields, including ecology, economics, social sciences, and computer science, among others.

At the heart of CAS is the concept of generative processes, which refers to the system's ability to create new patterns of behaviour and organisation through interactions among its components. In other words, a CAS's processes are generative processes of embodied interaction with the environment. More precisely, these interactions are not simply passive responses to environmental stimuli, but rather active engagements that shape and modify the environment itself (Kim and Park, 2013; Chemero, 2009; van Geert and de Ruiter, 2022; Hipólito, van Geert and Pessoa, 2023). The embodiment of CAS is a critical aspect of the system's dynamics. Agents are abstract entities that interact with the environment, but rather physical entities that possess agency and are capable of making decisions based on their perceptions of the environment. This embodiment allows agents to be situated within the environment, and to perceive, interpret, and act upon environmental stimuli in contextually appropriate ways. As CAS interact with their environment, they generate complex feedback loops that can lead to self-organisation and emergent behaviours. These feedback loops occur when the actions of one agent affect the environment, which in turn affects the actions of other agents. This can result in the emergence of patterns of behaviour and organisation that were not present in the system's initial state.

Understanding and modelling CAS is a challenging task because of the system's inherent complexity. A large number of interacting agents and the nonlinear dynamics of their interactions make it difficult to capture the full range of possible behaviours and outcomes. Even small changes in the initial conditions or parameters of the system can lead to significant changes in its behaviour and outcomes, making it challenging to predict the system's behaviour over time. Furthermore, because CAS is characterised by emergent behaviours and self-organisation, it is often impossible to predict these behaviours based solely on the properties of the individual components or the rules governing their interactions. Emergent behaviours and self-organisation arise from the interactions among agents and their environment, which are often context-dependent and influenced by factors that are difficult to observe or measure.

CASs processes are generative processes of embodied interaction with the environment, while generative models refer to mathematical models that capture the statistical regularities in the environment that give rise to observed behaviour emerging from embodied interaction. The main difference between the two is that generative model are typically formulated in terms of probabilistic relationships between hidden causes and observed sensory data, while generative processes refer to the actual behaving living system. While generative models are useful for making predictions and inferring the causes of sensory data, they do not necessarily capture the full complexity of the underlying biological processes that generate the sensory data. Because of the lack of computational tractability power, they employ dimensionality reduction a technique used in machine learning to reduce the number of variables in a dataset while still retaining as much information as possible. This is often done to simplify models and make them more manageable, as models with fewer variables are less complex and easier to work with. Generative models are used as a simplification of the more complex generative processes that are thought to underlie perception and action. In other words, while generative models are a useful tool for making predictions and understanding the statistical structure of the environment because they differ in levels of complexity, they are not the same as the actual bio-behavioural processes.

Adaptability and not low-dimension models of it, where complexity and, thereby, uncertainty, emerges Here lies the main issue in studying and developing real-world solutions for the optimisation of CAS, and specifically how to measure uncertainty in CAS. Even if generative models and processes can be correlated because they entail different levels of complexity, they are irreducible to one another. This paper develops a framework for the optimisation of CAS adaptation and attunement to their environment. Drawing from the Free Energy Principle, we stipulate that to remain alive a system must interact via purposefully context-specific actions that optimise environmentally adjusted development. The paper then begins by explaining that this tendency towards stability (i.e. seek preferred states in state space) can be seen as minimising free energy through active inference. It then theories, that, however, within an interaction's dynamical geometry high variability or noise, and over-specialized rigidity can impede adaptation and lead to a "stuck state" (i.e. a situation where a complex adaptive system (CAS) becomes overly specialised and rigid, hindering its ability to adapt to changing circumstances. The system's policy is optimised for a specific context but is not flexible enough to adapt to changes, leading to a state of low variability and high predictability. It then introduces the perspective that without perturbation (increasing the free energy), the system remains in the stuck state indefinitely. Finally, an important question is how much and what type of perturbation is required to move a CAS away from the minimum stuck state, which must be defined for a specific scale and the biological, psychological, and social factors unique to a Complex Adaptive System (CAS). We employ the construct of 'psychotic' Markov blankets can help understand and specify the variables and factors contributing to maladjustment conditions and offer treatment directions for the optimisation of living systems' adaptation and attunement to their environment from the neurobiological to psychological scales.

# 1. Modelling Embodied Attunement for Optimisation for Contextualised Action

Adapting means dynamically adjusting in an attunement way to the environment. It is a trajectory of being where a living system is able to dynamically interact with and adapt to its environment through optimised actions that are contextualised to the specific situation. Embodied attunement involves a reciprocal relationship between the body and the environment, where the body is attuned to the environment and the environment is attuned to the body. This means that our embodied interactions with the environment are not solely based on our sensory experiences, but also on our emotional experiences and our bodily movements and sensations. The embodiment of agents within Complex Adaptive Systems (CAS) is a critical aspect of the system's dynamics, as it enables physical entities to interact with their environment and make decisions based on their perceptions. This embodiment allows agents to be situated within the environment and interpret and act upon environmental stimuli in contextually appropriate ways. As CAS interact with their environment, they generate complex feedback loops that can lead to emergent behaviours and self-organization. These feedback loops occur when the actions of one agent affect the environment, which, in turn, affects the actions of other agents. The emergence of self-organization and the creation of emergent properties in CAS are a result of the interactions between agents and their environment, which lead to the emergence of patterns of behaviour and organization that were not present in the system's initial state. Therefore, it is crucial to consider the embodiment of agents within CAS to better understand their dynamics and the emergence of complex behaviours and organizations in such systems.

The concept of multistable interaction is central to the understanding of attunement as it refers to the ability of a living system to maintain multiple stable states of being in response to different environmental conditions (Pisarchik and Hramov, 2022). This means that the system is not rigidly locked into a single state but can adapt and shift between states as needed: context-specific actions that are optimised for achieving a specific goal to maintain health and well-being by adapting to the ever-changing environmental demands. This perspective emphasises the importance of flexibility, adaptability, and context-specificity in achieving and maintaining a state of health. It suggests that the key to health is not a fixed or static state, but rather a dynamic and adaptable process that involves ongoing interaction and optimisation with the environment – a form of embodied attunement. Context-specific actions are optimised for achieving a specific goal to adjust and attune to the environment in an optimal way to a context-specific situation. Adapting to the ever-changing environmental demands can be described as a tendency towards a stable point, where an agent strives to reach an optimal state that is conducive to its survival and flourishing.

Drawing from complex and dynamical systems theory, living systems can be understood in terms of trajectory, attractors, and repellors. Trajectory refers to the path that a living system takes through its development over time. Attractors and repellors, on the other hand, are states of the system that represent stable or unstable patterns of behaviour, respectively (Westley et al., 2013; Guckenheimer and Holmes, 2015; Kappel and Helbing, 2019; Levin, 2019; Prokopenko and Ay, 2021). In the context of a CAS development, attractors can be thought of as the different states that the system tends to settle into as it matures. For example, crawling, walking, and running are different attractor states for the locomotion of an animal. These attractor states are stable, meaning that the system tends to remain in them once it has achieved them (Thelen and Smith, 2003; Witherington and Boom, 2019; Iverson, 2021).

The tendency of systems to gravitate towards stable attractors can be explained by the principle of minimising free energy. Free energy is the difference between a system's predicted state and its actual state. When a system perceives its environment as uncertain or unpredictable, free energy increases, indicating a mismatch between the predicted and actual states of the system. At the level of individual agents, this translates to interactions with the environment to seek states that maintain their integrity. These agents are open systems that adjust and are attuned to their surroundings. Adaptive behaviour can be understood as active inference, where agents select actions that are most likely to lead to preferred outcomes while minimising the cost or surprise associated with sensory inputs. According to the FEP, living systems can be viewed as information-processing systems that strive to minimise their free energy. Free energy can be understood as the difference between the sensory information received from the environment and the predictions made by the internal models of the agent. The goal of the agent is to reduce this difference by acting upon the environment to make it more consistent with its internal models, which results in a reduction of the free energy. This process of minimising free energy can be understood in the context of active inference, where an agent selects actions that will reduce the discrepancy between its internal model and the sensory input from the environment. Active inference involves the integration of sensory information and prior beliefs to make predictions about the future and select actions that are expected to bring the sensory input closer to the predicted outcome.

Repellers, on the other hand, represent unstable patterns of behaviour. They are points in the system's trajectory that the system tends to avoid because they lead to undesirable outcomes. For example, if an animal has a tendency to fall over when it tries to walk too fast, then walking too fast would be a repeller for that animal. In this case, the animal's trajectory would tend to steer it away from the walking-too-fast state, and towards a more stable attractor state, such as walking at a moderate pace.

For adaptation, it is important to have a good balance between attractors and repellers. Attractions and repellers are not objectively "positive" or "negative", but it is the networked interaction between them that is relevant to understand the behaviour of a system in general, and to conjecture about its adaptive or adequate interaction with the environment. Driving these dynamics is purposeful action for the maintenance of the system. If for some reason, the system does not purposefully optimise context-specific action for its maintenance (minimise free energy), there is a risk that its trajectory will develop into a stuck state. A stuck state is a state where action optimisation is maximised. While specialisation can be useful, in a stuck state the system is not flexible to adapt to changes.

To adapt to the environment, it's important for a system to balance its attractions and repulsions. These forces aren't inherently "good" or "bad," but rather, it's the interaction between them that determines a system's behaviour and ability to adapt. The purpose of this interaction is to maintain the system, and if the system fails to do so by purposefully optimizing its actions for its environment, it risks becoming stuck in a fixed state. Specialization can be helpful, but in overspecialisation, the system loses its ability to adapt to changes. A stuck state is a state in which a system's behaviour becomes rigid and inflexible, limiting its ability to adapt to new or changing situations. In this state, the system has maximized its optimization over-specialized rigidity can hinder adaptation for its maintenance and is no longer flexible to adapt to changes.

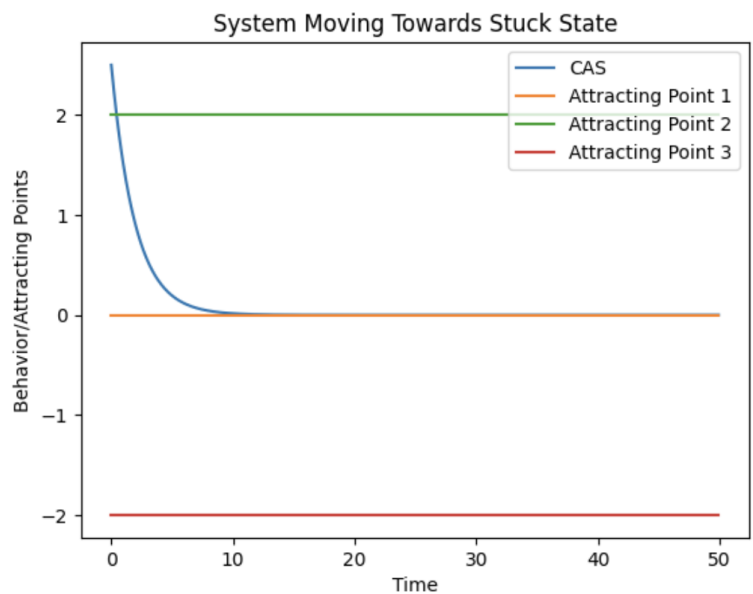

**Figure 1.** This time series that represents a Complex Adaptive System (CAS) moving towards three attracting points (y1, y2, and y3) over time. The system dynamics are defined by a simple linear function that calculates the difference between the current behaviour and the nearest attracting point. The plot shows the behaviour initially far from the attracting points and moving towards the closest one, but eventually getting stuck in a local minimum around one of the attracting points, indicating repetitive or rigid behaviour. The plot helps visualize how an individual's behaviour may be influenced by their environment and the tendency to seek stability in complex environments.

Behaviour can tend to converge towards the attracting point, which represents a stable pattern of behaviour that minimises the CAS uncertainty. This stable pattern is maintained for a period of time, indicating that the CAS has developed a policy that is optimised for this particular context. While optimisation is a natural tendency of purposeful action, over-optimisation Specialization can be advantageous, but when taken to the extreme, it can lead to a lack of adaptability within a system. This can result in a "stuck state" where the system's behaviour becomes rigid and inflexible, limiting its ability to adjust to new or changing situations (fig 2)

However, if this state becomes too stable it will become a stuck state, which is a state that will not be conducive to developing towards other possible stable points (fig. 2).

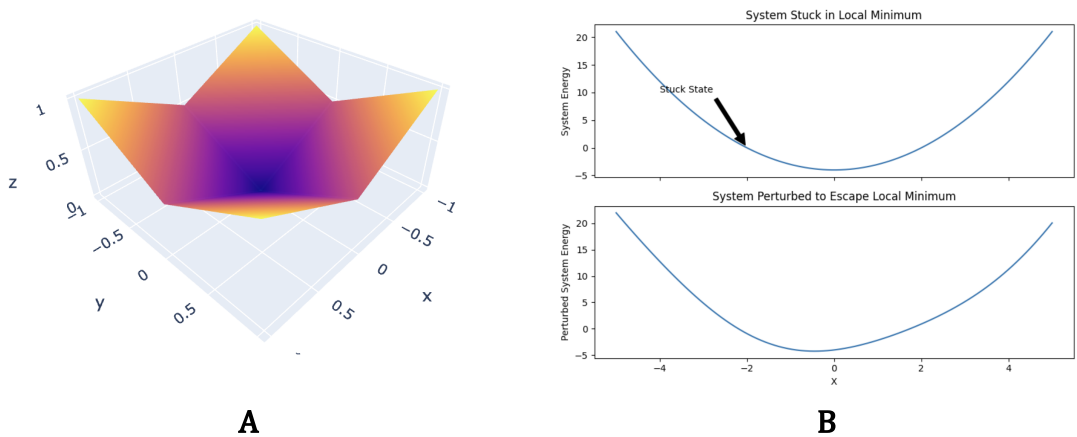

A                                            B

**Figure 2**, **A,** the surface plot diagram depicts two variables on the x and y axes, while the z-axis represents the "free energy" of the system. A "stuck state" occurs when the system reaches a minimum level of free energy. In this case, the minimum point of free energy is at (0,0), which corresponds to the bottom of the "bowl" in the surface plot. **B.** If the system is in a "stuck state," it means that it is located at or close to this minimum point and is incapable of moving away from it independently.

To facilitate the release of a system from a stuck state, a perturbation can be applied to increase the free energy and shift the system away from the minimum point. As illustrated in Figure 2, B, the introduction of a perturbation causes the system to move away from the minimum point of free energy and explore other regions of the surface plot. In this state, the system has optimized its performance, and overspecialization can impede adaptation by maintaining inflexibility and rigidity. Thus, it is crucial to strike a balance between specialization and flexibility to ensure a system's ability to adapt to changing circumstances.

Thus, it is crucial to consider the robustness of a CAS's policy when designing interventions or treatments aimed at supporting their behaviour. In Figure 3, when a high level of perturbation is introduced to the system at times 20 and 40, the CAS's behaviour deviates significantly from the attracting point and moves towards other attracting points, highlighting the importance of flexibility in behaviour and policy.

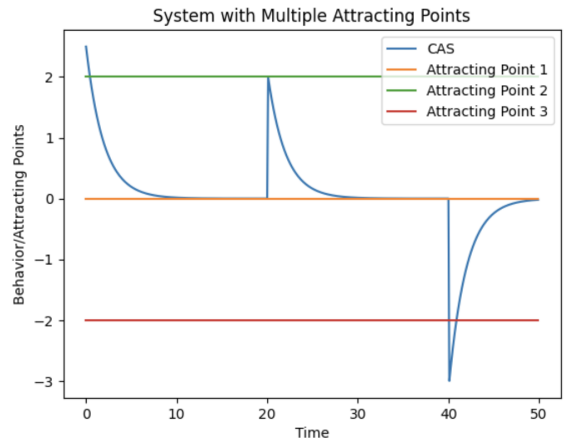

**Figure 3.** In Figure 3. The time series plot describes the behaviour of a system exhibiting multistability. Multistability refers to a system having multiple stable equilibrium points, also known as attracting points, that the system can settle into. In the case of the CAS depicted in the plot, there are three stable states that the system can settle into, depending on the initial conditions and any perturbations applied. At times 20 and 40, perturbations are introduced, causing the system to move away from the starting points.

Consequently, the use of stable points for maintaining rigid or repetitive behaviour can be disrupted (by increasing free energy) to bring about therapeutic change, as the system moves from a local minimum to a new stable state with improved performance. Nonetheless, although stable points may provide a sense of security, they can also constrain an individual's capacity to adapt to new or evolving circumstances. An intuitive example is this rigidity that makes it difficult for people with general mental health conditions, such as Autism Spectrum Conditions to participate in new experiences or interact with others in a dynamic social setting (for detail see Hipólito and White, 2023).

A critical open question pertains to the amount and type of perturbation required to break free from the stuck state, which must be defined for a specific scale and the biological, psychological, and social factors unique to a Complex Adaptive System (CAS). The subsequent section proposes the usefulness of the concept of 'psychotic' Markov blankets to comprehend and specify the variables and factors contributing to the maladjustment condition. Markov blankets to understand and specify factors contributing to maladjustment conditions, and discusses the importance of defining the amount and type of perturbation required to move a CAS away from the minimum stuck state. As Markov blankets are scale-free, they can define the quantity and nature of perturbation required for personalized intervention, therefore affording therapeutic and clinical application. The following sections then apply this framework to the nested neurobiological and psychological scales of life.

## 2. Perturbing Stuck States via Psychotic[1] Markov Blankets

It is worth noting that when no perturbation is introduced, the system remains indefinitely in a state of rigidity, also known as a "stuck state." This indicates that the individual's policy has been optimised for a specific context but lacks the flexibility to adjust to changes in the environment. A stuck state can be further described through the formalism of Markov blankets, what we shall call a 'psychotic' Markov blanket.

The Markov blanket is a statistical tool used to define a system's boundaries through conditional dependence or independence relationships, applicable to any self-organising system. It allows for the modelling of dependencies and dynamics between a system and its environment, emphasising the importance of understanding the interdependent relationship between them. The Markov blanket provides a framework for understanding the role of the environment in shaping a system's behaviour while maintaining its autonomy, by defining a set of variables that surround the system's internal states and labelling all other external variables. This set of variables is determined such that the internal states become conditionally independent from the external variables.

b = Markov blanket of μ

Mathematically speaking, the following equation defines the set of variables 'b' that make the internal states 'μ' conditionally independent from all other external variables 'η'. Simply put, if one knows the values of 'b', then predicting the behaviour of 'μ' becomes possi e, and no additional information from 'η' would be necessary to improve this prediction. Therefore, the equation can be written as:

$$p(\mu \mid b, \eta) = p(\mu \mid b) \qquad (1$$

The equation (1) includes the conditional probability distribution of 'μ' given both 'b' and 'η', which is represented as p(μ | b, η), and the conditional probability distribution of 'μ' given only 'b', which is represented as p(μ | b).

---

[1] The term 'psychotic' is often used in clinical settings to describe a state of mind that is characterised by a significant departure from reality. However, it is important to note that this term is not meant to be taken literally, but rather as a means of conveying the idea of a loss of touch with reality. The use of this term is intended to capture the sense of disconnection and insulation from the external environment that can be observed in individuals who are experiencing significant mental health challenges.

In a dynamic system, equation (1) indicates that the average rate of change of each component in a Markov blanketed system can be influenced by just two other types of states to maintain the equation's underlying structure.

$$\dot{\mu} = f_\mu(\mu, s, a)$$
$$\dot{a} = f_a(\mu, s, a)$$
$$\dot{\eta} = f_\eta(\eta, s, a)$$
$$\dot{s} = f_s(\eta, s, a) \tag{2}$$

Equation 2 states that the internal and active states of a system are dependent on the blanket states, which include internal, sensory, and active states. Similarly, the external and sensory states of a system are dependent on the blanket states, which include external, sensory, and active states. This implies that the current state of a system is determined by the interactive dynamics between internal, sensory, and active states, while the current state of the environment is determined by the dynamics between external, sensory, and active states. Consequently, the internal and external states of a system indirectly influence each other in a reciprocal manner. Figure **2A** illustrates this reciprocal influence.

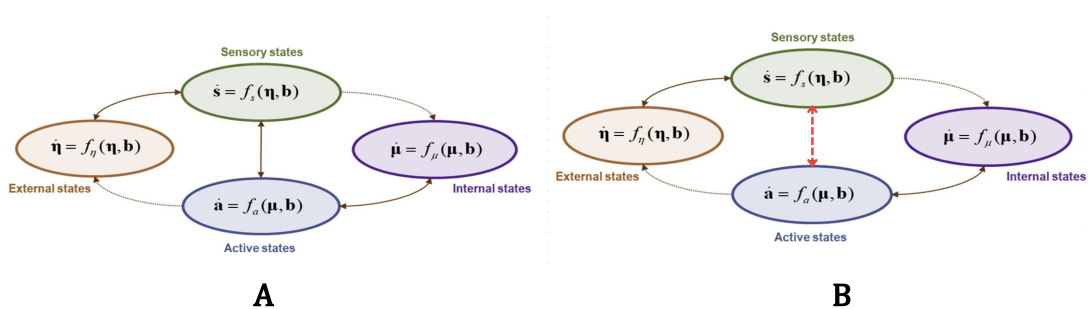

A     B

**Figure 3 A.** Internal (purple) and external (orange) states are conditionally independent of one another. Mathematically, this means that they reciprocally influence each other by the dynamical influences within blanket states, i.e. sensory (green) and active (blue) states. A balanced influence between blanket states translated into adapted behaviour. **B.** Should the reciprocal influences within blanket states be broken (red dotted arrow), then internal states will be insulated, i.e. they will not be influenced by the environment (external states) for flexibility in adapting to changing circumstances.

Active inference is a theoretical framework that aims to explain how biological systems maintain their internal states by minimizing prediction errors or surprise. To illustrate, consider a Complex Adaptive System (CAS) with a single internal state variable whose goal is to avoid a fixed point in state space. By using active inference, the CAS minimizes its prediction error and maintains its internal state away from the fixed point. This behaviour can be modelled as a Markov decision process (MDP), where the behaviour at each time step depends on the current state of the system and a policy that maps the state to action. The goal is to minimize free energy, which measures the difference between actual and expected sensory input.

Purposeful contextual sensitive action is taken to minimise uncertainty. This can be modelled by a Markov Decision Process MDP consisting of a set of states and actions, with probabilities t t describe the transition from one state to another based on the policy. By choosing an action that moves the system towards one of the attracting points, the policy reduces the uncertainty about the environment and thus minimises free energy. The policy can be optimised using techniques from reinforcement learning, which involves learning a mapping from states to actions that maximise a reward signal.

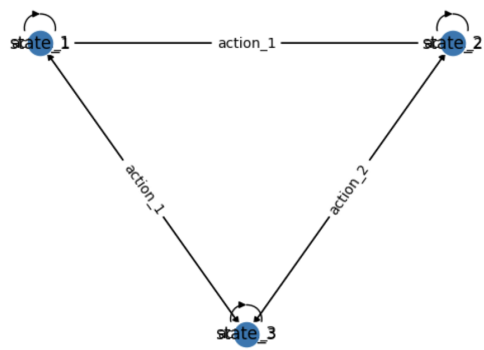

**Figure 5.** The diagram depicts Markov Decision Process (MDP) with three states and three actions. The states represent stable equilibrium points, and actions represent attracting points for autistic people to reduce uncertainty about their environment and minimize free energy. At each time step, the system is in one of the three states, and the CAS can take one of the three actions. The transition probabilities indicate the likelihood of moving from one state to another when taking a specific action. For example, taking action 1 while in state 1 has a 0.9 probability of remaining in state 1 and a 0.05 probability of moving to state 2 or 3.

An MDP is a stochastic control process that consists of a set of states, a set of actions, and a set of probabilities that describe the transition from one state to another, given a particular action. The probabilities that describe the transition from one state to another are determined by the policy that maps the state to action. The policy can be seen as a way to optimise the behaviour of the system in order to minimise free energy.

However, if the internal state of the CAS fails to move towards the prior mean over time (fig 3 B) by engaging in activities that affords the minimisation of uncertainty, it could imply that the organism is not adapting to changes in the environment or is not sustaining a healthy relationship with it. This suggests that the organism's beliefs about the environment are not being updated effectively to enable adaptation to changes in the environment. This can have negative consequences, such as the inability to cope with environmental changes or failure to maintain a sustainable relationship with the environment.

In conclusion, the system is insulated in a stuck state. 'Psychotic' Markov blanket provides a formal way to define the states and factors contributing to the stuck state and understand how an individual's cognitive processes and behaviours are influenced by their interaction with the environment. Interpreting a "stuck state" in mental health conditions as a Markov blanket unbalances between sensory and active states can provide insights into the underlying mechanisms of the condition and guide targeted interventions to restore balance and promote recovery. Specifically, identifying the amount and type of perturbation necessary for moving a CAS away from a stuck state towards multistability or multistable interaction of embodied attunement. Further, the next section proposes that the insulation in a stuck can be overcome by applying a perturbation to the Markov blanket. In other terms, by increasing the free energy in the system. This should move a CAS away from its dynamics conforming with those described as a 'psychotic' Markov blanket, thereby away from a stuck state.

To sum up, over-specialized rigidity can lead to a "stuck state" which can be understood through the construct of a "psychotic" Markov blanket. Interventions can help a system overcome the stuck state and move towards multistability by applying sufficient perturbation. Such interventions can focus on altering attractors or repellers in the system or promoting greater adaptability to the environment. The construct of a "psychotic" Markov blanket offers a formal definition of the factors contributing to the stuck state, and understanding these factors can guide targeted interventions for promoting recovery by restoring balance. Identifying the amount and type of perturbation necessary for moving a complex adaptive system (CAS) away from a stuck state towards multistability is crucial. The next section proposes that increasing free energy in the system through perturbation can overcome the insulation in a stuck state associated with a "psychotic" Markov blanket. This can help CAS develop more adaptive patterns of behaviour, cognition, and emotion, and navigate their environment with greater ease as we shall see in the next sections, from microscale cellular levels to behaviour.

3. Cellular CAS

At the **cellular level,** living organisms face the challenge of maintaining a stable internal environment despite external environmental fluctuations. This is accomplished through homeostatic mechanisms that serve to minimise free energy by preserving a constant internal environment.

For example, temperature regulation, osmoregulation, and pH regulation are all mechanisms that help to maintain cellular stability. These mechanisms can be seen as context-specific actions that are optimised to sustain cellular health and well-being by adapting to the changing environmental demands. The active inference process allows the cell to act upon the world to achieve preferred states, minimising free energy and maintaining stability. This highlights the importance of context-specific actions and the need for adaptation to achieve and maintain health at the cellular level. At the cellular level, the actions that are involved in maintaining a stable internal environment may include the uptake and excretion of certain molecules, regulation of metabolic processes, and communication with neighbouring cells. For example, cells may actively pump ions across their membrane to regulate the concentration of ions in the cytoplasm, which is important for maintaining proper cellular function.

In the brain, systems and network neuroscience show that neural activity is highly integrated (Sporns, 2013; Wasserman and Wasserman, 2023), the concept of Markov blankets is an important aspect of systems neuroscience and provides a framework for understanding the interactions between different levels of organisation in the brain. Markov blankets in the brain demarcate boundaries of couplings from pairs of neurons to cortical columns and brain-wide networks. The presence of Markov blankets in the brain enables partitions into single neurons, brain regions, and brain-wide networks, which can be used to study the connectivity and function of the brain at multiple scales (Hipólito et al., 2021; Friston et al., 2021).

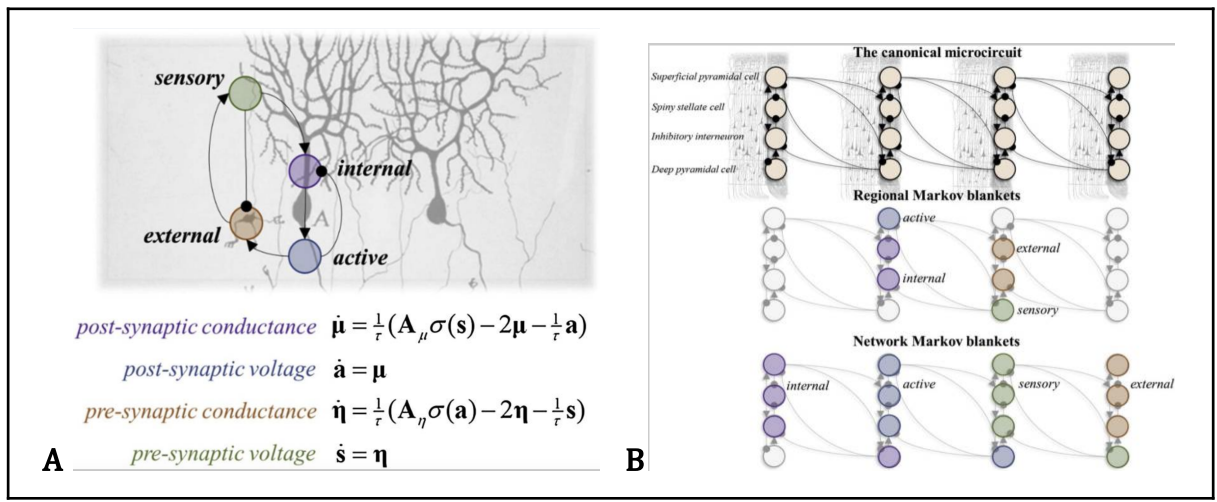

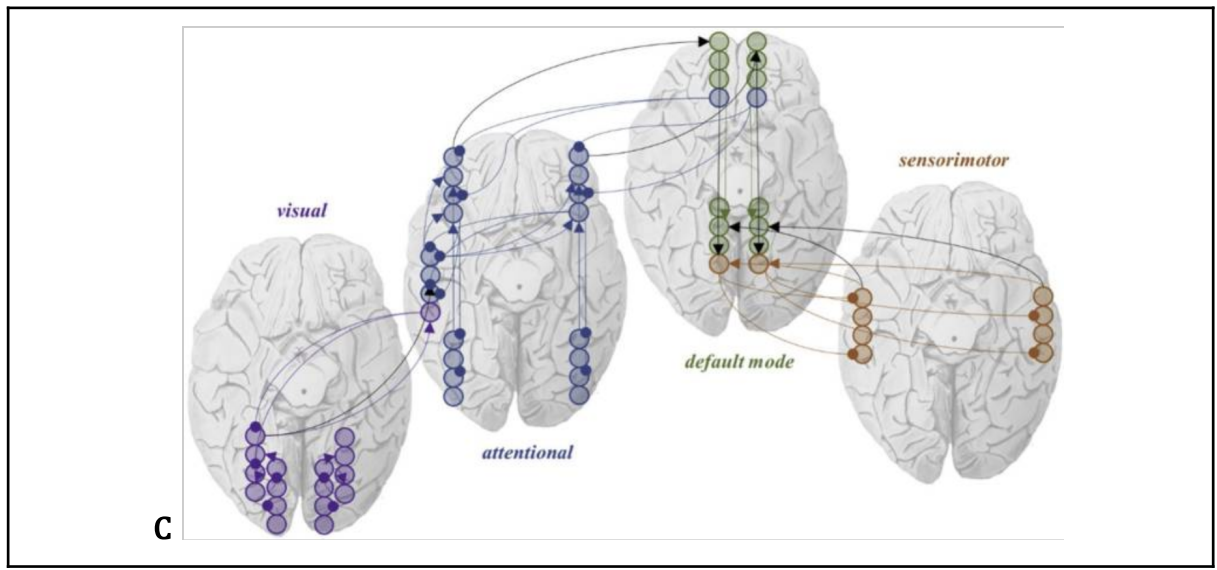

**Figure 4. A.** The Neuronal Markov blankets figure shows a pair of neurons separated by a Markov blanket, with constants A acting as connectivity strengths from the active state of one neuron to the external state of the other and from the sensory states of the latter to the internal states of the former. The sigma function converts potentials to firing rates. The structure has a unique feature whereby the sensory states can arise from many external states while the active states depend only on the conductance of the neuron being depolarized. The figure also distinguishes between excitatory and inhibitory influences using different arrowhead shapes. **B.** This section describes cortical micro-circuitry and its representation using Markov blankets. The upper schematic shows the connectivity of the canonical microcircuit consisting of four cell populations with a specific pattern of connectivity. The second row illustrates the Markov blankets that underlie the separation into distinct cortical regions. The final row shows a separation into a network of regions, with the middle two regions acting to insulate the far left and right regions. The dynamics of each neural population obey the equations given in Fig. 3A, where the likelihood mappings specify which populations are connected to one another. Feedforward connections originate predominantly from superficial layers, and feedback connections from deep layers. **C.** The image in this figure represents a Markov blanket of networks, where the connections between nodes in different networks are treated as dependencies between states. The networks themselves act as the active, sensory, internal, and external states, loosely structured around resting-state fMRI studies. The visual networks are treated as internal states that influence active states, while the default mode network represents sensory states mediating the influence between internal and external states. The assignment of these states is equally valid if reversed. For detail see Hipólito et al., 2021).

Markov blankets can be useful in understanding the complex interactions between different parts of the brain, such as in the case of the relationship between the prefrontal cortex and the basal ganglia in decision-making. By identifying the Markov blanket of each region, researchers can create more accurate models of how these regions interact and influence each other.

Conversely, evidence shows that occurrences of lack of communication between neural cells may be associated with psychopathological conditions (Kaiser et al., 2015; van den Heuvel and Sporns, 2019), such as schizophrenia, which is known as the Dysconnection Hypthesis (Friston et al., 2016; Limong et al., 2023; Sapienza et al, 2023). The dysconnection hypothesis proposes a failure of functional integration in neuronal systems dependent on long-range connections, isolating abnormalities in brain function: the underlying cause of dysconnectivity in schizophrenia is a specific impairment of

synaptic plasticity, which results from aberrant modulation of NMDAR function by DA, ACh, and 5-HT

Perturbation to move a CAS away from a stuck state (i.e. dysconnection of a 'psychotic' Markov blanket) for multistable integration could involve various methods such as (1) neurofeedback training, brain stimulation techniques (e.g. transcranial magnetic stimulation or transcranial direct current stimulation), or pharmacological interventions that target specific neurotransmitter systems known to affect functional connectivity; (2) Brain stimulation techniques can be targeted to specific regions or networks in the brain to enhance functional connectivity and reduce dysconnection. (3) Pharmacological interventions may also be used to enhance functional connectivity by targeting specific neurotransmitter systems. For example, drugs that modulate the levels of dopamine, glutamate, or GABA in the brain have been shown to affect functional connectivity and may be useful in treating dysconnection syndromes.

By learning to modulate their brain activity, individuals may be able to improve functional integration and reduce dysconnection. Models can further investigate how multidimensional landscape determines the fate of a cell (Sáez, Briscoe and Rand, 2022; as well as applying complexity-inspired frameworks, (1) complexity, (2) criticality, (3) controllability, and (4) coordination to analyse resting-state neuroimaging data and how they can provide insight into the organisation and dynamics of brain activity (Hancock et al., 2022).

## 4. Mental health and well-being as a CAS

Studying mental health as a CAS, not as symptoms, mental health can be conceptualised as complex biopsychosocial systems that can tend towards a "stuck state". These conditions affect a person's thinking, feeling, behaviour, or mood and deeply impact day-to-day living, often affecting their ability to relate to others. However, it's important to note that mental health problems are not categorical idealizations, but rather complex biopsychosocial processes that unfold in individuals over time (Cramer et al., 2010; McNally, 2016; Fried et al., 2017; Cacioppo and Cacioppo, 2018; Roefs et al., 2022; Bringmann et al., 2023).

Mental health as a CAS involves recognizing that mental health conditions are a product of a complex biopsychosocial system that includes both bottom-up (genetic and endocrinological) and top-down (social and psychological) influences. When this system is disrupted by adverse events, such as trauma or chronic stress, it can result in a state of "stuckness" where the individual's ability to adapt and recover is compromised.

When an individual is in a "stuck state", it can be interpreted as an unbalance in the Markov blanket between sensory and active states. Sensory states involve the individual's perception and processing of external stimuli, while active states involve the individual's ability to take action and respond to the environment. An unbalance between these two states can result in a situation where the individual is unable to respond appropriately to external stimuli, leading to a "stuck state". For example, in individuals with post-traumatic stress disorder (PTSD), the Markov blanket may be unbalanced towards sensory states, where they may experience hyperarousal and hypervigilance to potential threats. This unbalance may result in the individual being unable to differentiate between real and perceived threats, leading to a "stuck state" where they are unable to respond appropriately to their environment.

A psychotherapeutic intervention that aims to perturb a mental condition as a stuck state has the overarching goal of promoting recovery and restoring balance to the individual's mental health. The specific aim of this type of intervention is to disrupt the individual's existing patterns of thought and behaviour that may be contributing to their stuck state and encourage adaptation and growth. In a mental health condition, an individual feels trapped or unable to make progress towards their goals or desires. This can manifest in a variety of ways, such as feeling constantly sad or anxious, experiencing intrusive thoughts or memories, or feeling disconnected from one's environment or loved ones. A psychotherapeutic intervention that aims to perturb this stuck state seeks to help the individual break free from these patterns and develop new ways of thinking and behaving that are more adaptive and supportive of their mental health.

The specific aims of a perturbation-based intervention will vary depending on the individual and their unique circumstances. For example, an intervention for depression may aim to increase the individual's sense of agency and control over their environment, while an intervention for anxiety may focus on reducing avoidance behaviours and increasing tolerance for uncertainty. The overall goal, however, is to create a perturbation or disruption that can help the individual move beyond their stuck state and towards a more adaptive and fulfilling way of life.

Ultimately, perturbation-based interventions must be carefully tailored to the individual's needs and abilities. The goal is to create a challenge that is manageable and supportive of the individual's growth, rather than overwhelming or triggering. A skilled psychotherapist can help guide the individual through this process and provide support and guidance as they navigate the challenges of perturbation-based interventions.

## Conclusion

In conclusion, this paper presents a novel framework for optimising the adaptation and attunement of complex adaptive systems (CAS) with their environments. The paper highlights the tendency towards stability in CAS through active inference, but also identifies the risks of over-specialization and rigidity leading to a "stuck state." To address this, the paper introduces the concept of 'psychotic' Markov blankets and emphasises the importance of perturbation to increase free energy and enable CAS to adapt to changing circumstances. The paper provides directions for applying this framework to real-world problems at various levels of complexity, from cells to societies and ecosystems. Overall, this framework has the potential to guide targeted interventions and promote optimal adaptation and attunement in CAS across a wide range of domains.

### Glossary 1: Complex Systems Theory

**Attractors:** States towards which a system tends to move, and which it tends to remain in.
**Complex system:** a system composed of interconnected and interdependent parts that exhibit emergent behaviour and are difficult to predict or control.
**Emergence:** The appearance of new properties or patterns at higher levels of organization that cannot be explained solely by the properties or behaviours of the system's individual components.
**Feedback loops:** Mechanisms by which a system's outputs are fed back into its inputs, creating a cycle of cause and effect.
**Multistability:** The property of a system to have more than one stable state or attractor, which can result in alternative outcomes depending on initial conditions.
**Nonlinearity:** the property of a system where the relationship between cause and effect is not proportional or additive, making it difficult to predict the behaviour of the system.
**Repellers:** States away from which a system tends to move, and which it tends to avoid.
**Stable states:** States that a system can achieve and maintain, which can be either attractors or repellers.
**Trajectory:** the path that a system follows over time as it evolves and changes. This can refer to the position, state, or behaviour of individual components within the system, or to the overall behaviour of the system as a whole.

### Glossary 2: Free Energy Principle

> **Active inference:** The idea that organisms actively seek to minimize the free energy of their sensory states by making predictions about the world and adjusting their behaviour to match those predictions.
> **Free energy expectation:** The prediction generated by an organism's internal model of the world, which is compared to the actual sensory input it receives.
> **Free energy minimization:** The process by which an organism reduces the discrepancy between its internal model of the world and the sensory information it receives, in order to minimize the amount of free energy in its system.
> **Free Energy Principle:** The idea that organisms, including the brain, are driven by a fundamental imperative to minimize free energy in their internal states, by maintaining their internal models in alignment with the external world.
> **Free energy:** a measure of the amount of internal uncertainty or disorder in a system that a system seeks to minimize over time by converging towards stable patterns or attractors.
> **Generative models:** A mathematical framework that allows for the generation of predictions about the world based on internal models of the environment.
> **Generative processes:** The processes by which an organism generates predictions about the sensory input it expects to receive from the environment, based on its internal model of the world.
> **Markov blankets:** The boundary that separates a system from its environment, defining what is internal and what is external to the system.
> **Policy:** a set of rules or strategies that govern the behaviour of a system or agent in a particular context.

## References

Axelrod, R. (1997). The complexity of cooperation: Agent-based models of competition and collaboration. Princeton University Press.

Borsboom, D., & Cramer, A. O. (2013). Network analysis: An integrative approach to the structure of psychopathology. Annual Review of Clinical Psychology, 9, 91-121. https://doi.org/10.1146/annurev-clinpsy-050212-185608

Bringmann, L. F., & Eronen, M. I. (2018). Don't blame the model: Reconsidering the network approach to psychopathology. Psychological Review, 125(4), 606-615. https://doi.org/10.1037/rev0000103

Bringmann, L., Helmich, M., Eronen, M., & Voelkle, M. (2023). 5 Complex Systems Approaches to Psychopathology. *Oxford Textbook of Psychopathology*, 103.

Cacioppo, J. T., & Cacioppo, S. (2018). Studying mental health problems as systems, not symptoms. Clinical Psychological Science, 6(4), 529-542. https://doi.org/10.1177/2167702617745092

Chemero, A. (2009). Radical embodied cognitive science. MIT Press.

Cramer, A. O., Waldorp, L. J., van der Maas, H. L., & Borsboom, D. (2010). Comorbidity: A network perspective. Behavioral and Brain Sciences, 33(2-3), 137-150. https://doi.org/10.1017/S0140525X09991567

Del Ser, J., Osaba, E., Molina, D., Yang, X. S., Salcedo-Sanz, S., Camacho, D., ... & Herrera, F. (2019). Bio-inspired computation: Where we stand and what's next. *Swarm and Evolutionary Computation*, *48*, 220-250.

Fried, E. I., van Borkulo, C. D., Cramer, A. O., Boschloo, L., Schoevers, R. A., & Borsboom, D. (2017). Mental disorders as networks of problems: A review of recent insights. Social Psychiatry and Psychiatric Epidemiology, 52(1), 1-10. https://doi.org/10.1007/s00127-016-1319-z

Friston, K. J., Fagerholm, E. D., Zarghami, T. S., Parr, T., Hipólito, I., Magrou, L., & Razi, A. (2021). Parcels and particles: Markov blankets in the brain. *Network Neuroscience*, *5*(1), 211-251.

Friston, K., Brown, H. R., Siemerkus, J., & Stephan, K. E. (2016). The dysconnection hypothesis (2016). *Schizophrenia research*, *176*(2-3), 83-94.

Friston, Karl; Brown, Harriet R; Siemerkus, Jakob; Stephan, Klaas E (2016). *The dysconnection hypothesis (2016)*. Schizophrenia Research, 176(2-3):83-94.

Grigolini, P. (2015). Emergence of biological complexity: Criticality, renewal and memory. *Chaos, Solitons & Fractals*, *81*, 575-588.

Guckenheimer, J., & Holmes, P. (2015). Nonlinear oscillations, dynamical systems, and bifurcations of vector fields (Vol. 42). Springer.

Hancock, F., Rosas, F. E., Mediano, P. A., Luppi, A. I., Cabral, J., Dipasquale, O., & Turkheimer, F. E. (2022). May the 4C's be with you: an overview of complexity-inspired frameworks for analysing resting-state neuroimaging data. *Journal of the Royal Society Interface*, *19*(191), 20220214.

Hipólito, I., Ramstead, M. J., Convertino, L., Bhat, A., Friston, K., & Parr, T. (2021). Markov blankets in the brain. *Neuroscience & Biobehavioral Reviews*, *125*, 88-97.

Hipólito, I., van Geert, P., & Pessoa, L. (2023). Mind as Motion: Patterns from Psychology to Neurobiology.

Holland, J. H. (1995). Hidden order: How adaptation builds complexity. Addison-Wesley.

Iverson, J. M. (2021). Developmental variability and developmental cascades: Lessons from motor and language development in infancy. *Current Directions in Psychological Science*, *30*(3), 228-235.

Kaiser, R. H., Andrews-Hanna, J. R., Wager, T. D., & Pizzagalli, D. A. (2015). Large-scale brain networks and psychopathology: A unifying triple network model. Trends in Cognitive Sciences, 19(10), 534-550.

Kappel, D., & Helbing, D. (Eds.). (2019). Modeling Complex Living Systems: A Kinetic Theory and Stochastic Game Approach. Springer.

Kim, C. S. (2023). Free energy and inference in living systems. *Interface Focus*, *13*(3), 20220041.

Kim, J., & Park, J. (2013). An embodied cognition approach to the interactive generative animation. In Proceedings of the 19th ACM international conference on Multimedia (pp. 1025-1028).

Levin, S. A. (2019). Fragile dominion: Complexity and the commons. Basic Books.

Levin, S. A. (2019). Fragile dominion: Complexity and the commons. Basic Books.

Levinthal, D. A. (1997). Adaptation on rugged landscapes. Management Science, 43(7), 934-950. https://doi.org/10.1287/mnsc.43.7.934

Limongi, R., Jeon, P., Mackinley, M., Das, T., Dempster, K., Théberge, J., ... & Palaniyappan, L. (2020). Glutamate and dysconnection in the salience network: neurochemical, effective connectivity, and computational evidence in schizophrenia. *Biological psychiatry*, *88*(3), 273-281.

McNally, R. J. (2016). Can network analysis transform psychopathology?. Behaviour Research and Therapy, 86, 95-104. https://doi.org/10.1016/j.brat.2016.06.005

Pisarchik, A. N., & Hramov, A. E. (2022). *Multistability in Physical and Living Systems*. Berlin: Springer.

Prokopenko, M., & Ay, N. (Eds.). (2021). A General Theory of Complex Living Systems: Exploring the Demand Side of Dynamics. Springer.

Roefs, A., Fried, E. I., Kindt, M., Martijn, C., Elzinga, B., Evers, A. W., ... & Jansen, A. (2022). A new science of mental disorders: Using personalised, transdiagnostic, dynamical systems to understand, model, diagnose and treat psychopathology. *Behaviour Research and Therapy*, *153*, 104096.

Sáez, M., Briscoe, J., & Rand, D. A. (2022). Dynamical landscapes of cell fate decisions. Interface focus, 12(4), 20220002.

Sapienza, J., Bosia, M., Spangaro, M., Martini, F., Agostoni, G., Cuoco, F., ... & Cavallaro, R. (2023). Schizophrenia and psychedelic state: Dysconnection versus hyper-connection. A perspective on two different models of psychosis stemming from dysfunctional integration processes. *Molecular Psychiatry*, *28*(1), 59-67.

Smith, L. B., & Thelen, E. (2003). Development as a dynamic system. *Trends in cognitive sciences*, *7*(8), 343-348.

Sporns, O. (2013). The human connectome: Origins and challenges. NeuroImage, 80, 53-61.

Stephan, K. E., Friston, K. J., & Frith, C. D. (2009). Dysconnection in schizophrenia: from abnormal synaptic plasticity to failures of self-monitoring. *Schizophrenia bulletin*, *35*(3), 509-527.

Sterman, J. D. (2000). Business dynamics: Systems thinking and modeling for a complex world. Irwin/McGraw-Hill.

van den Heuvel, M. P., & Sporns, O. (2019). A cross-disorder connectome landscape of brain dysconnectivity. Nature Reviews Neuroscience, 20(7), 435-446. https://doi.org/10.1038/s41583-019-0175-4

van Geert, P., & de Ruiter, N. (2022). *Toward a Process Approach in Psychology: Stepping Into Heraclitus' River*. Cambridge University Press.

Wasserman, T., & Wasserman, L. D. (2023). The Human Connectome: An Overview. *Apraxia: The Neural Network Model*, 35-48.

Westley, F., Tjornbo, O., Schultz, L., Olsson, P., Folke, C., Crona, B., & Bodin, Ö. (2013). A theory of transformative agency in linked social-ecological systems. Ecology and Society, 18(3), 27. https://doi.org/10.5751/ES-05072-180327

Witherington, D. C., & Boom, J. (2019). Developmental dynamics: Past, present, and future. *Human Development*, *63*(3-4), 264-276.